\documentclass[twocolumn,showpacs, prb]{revtex4}
\usepackage{color}
\usepackage{graphicx}
\usepackage{dcolumn}
\usepackage{bm}

\begin{document}

\title{Armchair graphene nanoribbons: Electronic structure and electric field modulation}

\author{Hassan Raza and Edwin C. Kan}
 \address{School of Electrical and Computer Engineering, Cornell University Ithaca NY 14853 USA}%

\begin{abstract}
We report electronic structure and electric field modulation calculations in the width direction for armchair graphene nanoribbons (acGNRs) using a semi-empirical extended H\"uckel theory. Important band structure parameters are computed, \textit{e.g.} effectives masses, velocities and bandgaps. For the three types of acGNRs, the $p_z$ orbital tight-binding parameters are extracted if feasible. Furthermore, the effect of electric field in the width direction on acGNRs dispersion is explored. It is shown that for the two types of semiconducting acGNRs, an external electric field can reduce the bandgap to a few meV with different quantitative behavior. 
\end{abstract}



\maketitle

\begin{figure}
\vspace{2.0in}
\hskip-3.0in\includegraphics{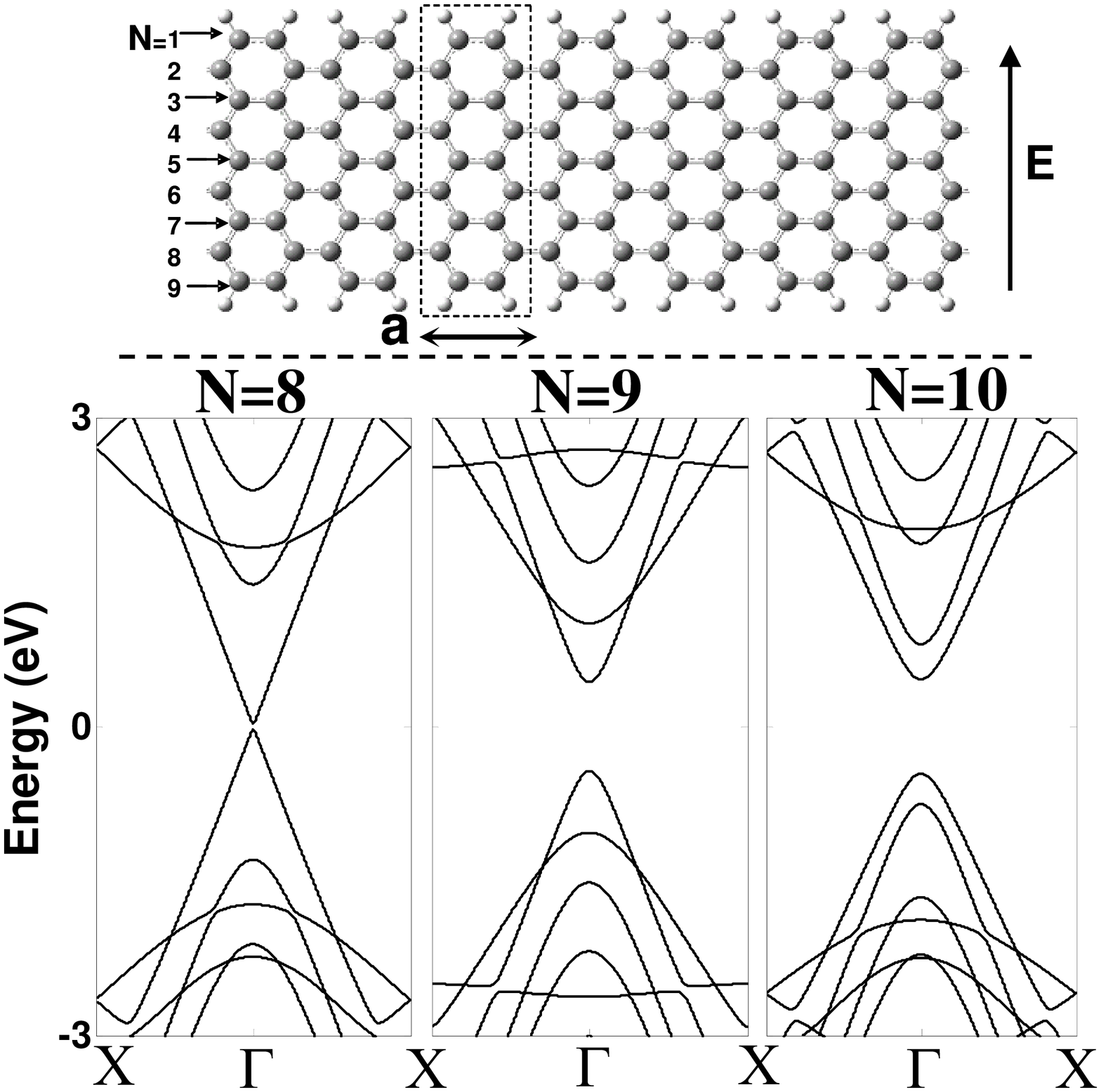}
\caption{Electronic structure of armchair graphene nanoribbons (acGNR). The ball and stick model of a graphene nanoribbon with N=9 is shown with the unit cell. E-k diagrams are shown for three different types of acGNRs using extended H\"uckel theory (EHT).}
\end{figure}

\section{Introduction} Unconstrained graphene is a two dimensional hexagonal monolayer of carbon atoms. Its unique linear dispersion around the Dirac point and zero bandgap \cite{Wallace47,Saito98} has generated significant interest \cite{Geim07, Neto08}. Constraining one dimension of graphene results into nanoribbons. The electronic structure of these graphene nanoribbons (GNR) depends on the width and chirality \cite{Nakada96, Kawai00, Raza08, Fujita96, Zhihong07, Brey06, Barone06, Wakabayashi99}. Two unique GNRs are armchair and zigzag referred to as acGNR and zzGNR in this article. acGNR has an armchair edge as shown in Fig. 1 and when conceptually rolled to form a nanotube results in a zigzag tube and \textit{vice versa}. Some experimental techniques have already been used to measure their properties \cite{Han07} and numerous fabrication schemes have been devised \cite{Li08, Novoselov04, Berger06}. Electronic applications of graphene and GNRs are also being sought after \cite{Fiori07, Liang07, Feenstra07}.

\begin{figure*}
\vspace{2.0in}
\hskip-3.0in\includegraphics{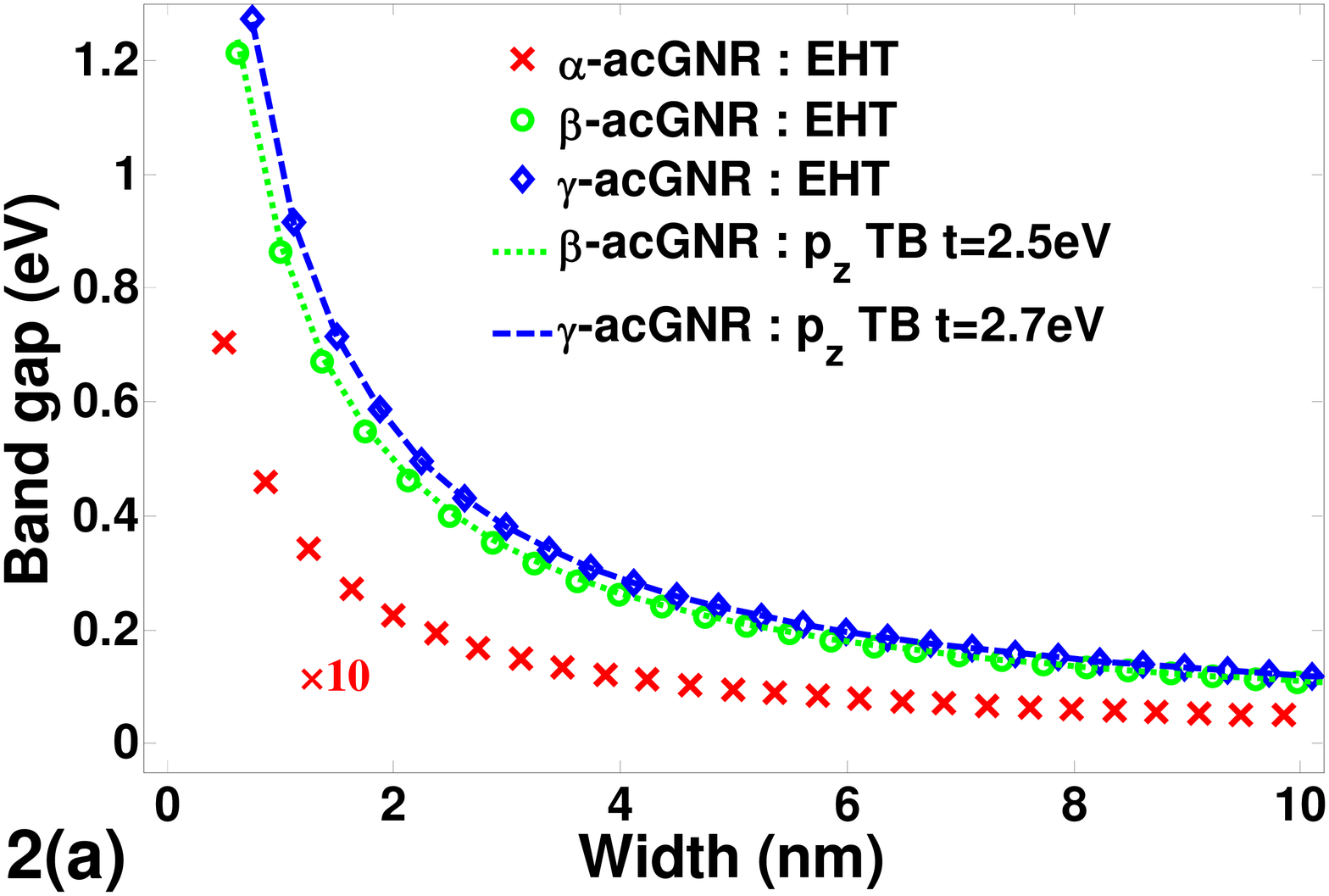}
\hskip3.6in\includegraphics{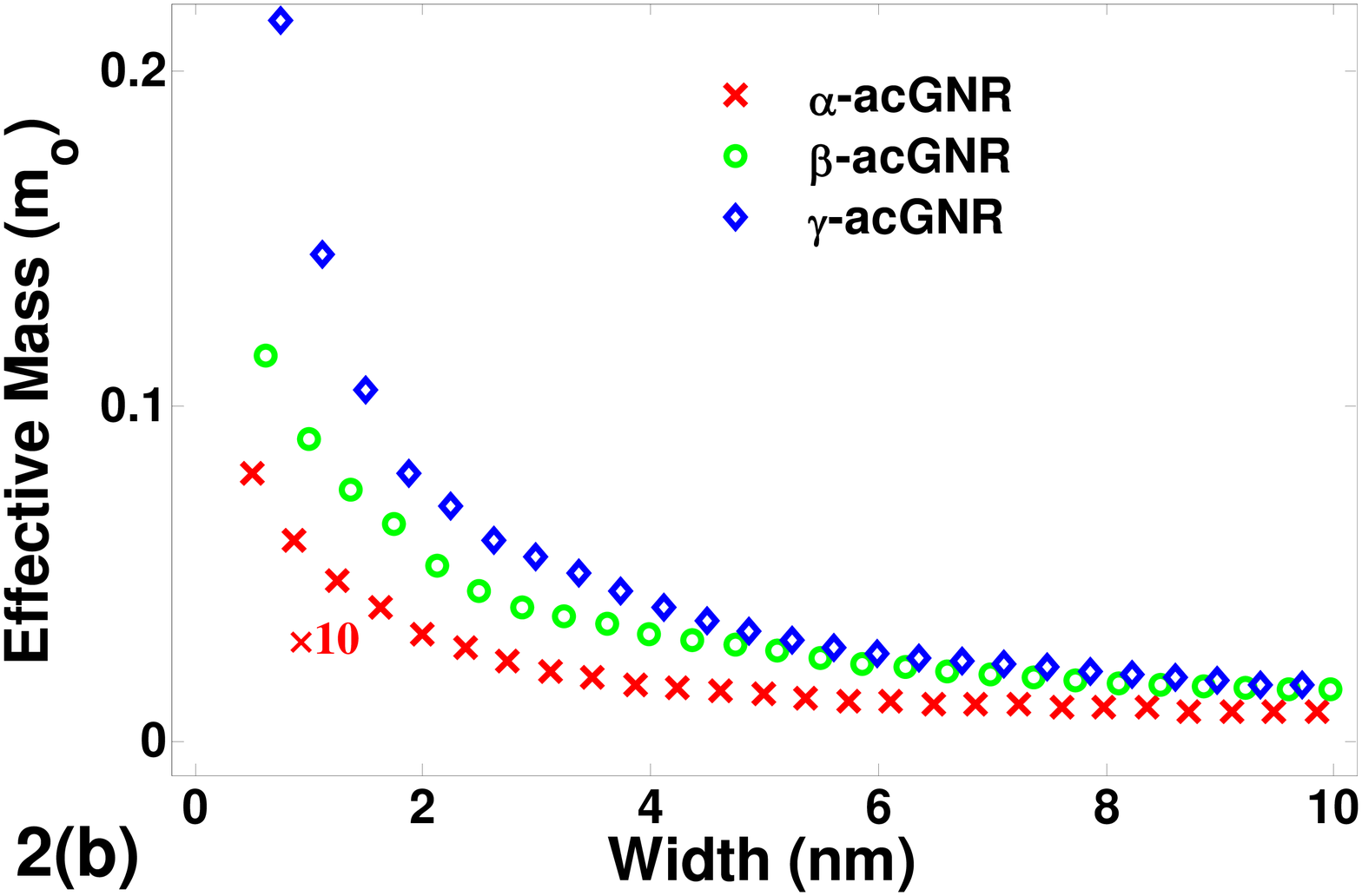}
\caption{(color online) Bandgaps and effective masses. (a) Variation of bandgap with nanoribbon widths of different types of acGNRs. Using a $p_z$-orbital tight binding method, t=2.5eV and t=2.7eV match the bandgaps obtained by extended H\"uckel theory (EHT) for $\beta$-acGNRs and $\gamma$-acGNRs, respectively. (b) Variation of effective mass with nanoribbon widths of different types of acGNRs.}
\end{figure*}

In zzGNRs, the wavefunctions for conduction and valence bands are localized at the edges \cite{Nakada96,Fujita96}. In addition, the bands around the Fermi energy have very small dispersion that leads to Stoner magnetism \cite{Nakada96,Fujita96}. These edge states can be modulated with an external electric field in the width direction, resulting in half metallicity \cite{Son06}. 

In acGNRs, the wavefunctions associated with bands around Fermi energy are distributed throughout the width of the nanoribbon. However, these bands still can be modulated with an external electric field in the width direction as discussed by Novikov using a continuum model \cite{Novikov07}. In addition, due to quantization in one direction, acGNRs have velocities less than those found in unconstrained graphene sheets, and the band structure has a parabolic character around the band edge within a few tens of meV. 

In this paper, we focus on acGNRs and study their electronic structure and electric field modulation in the width direction with a semi-empirical extended H\"uckel theory (EHT). Similar electric field modulation effects have been studied in carbon nanotubes as well \cite{Li02, Novikov06}. The detailed model has been reported in Ref. \cite{Raza08}. EHT parameters are transferable and have been benchmarked with generalized gradient approximation of density functional theory (DFT) for carbon atoms in graphene structure. EHT is computationally inexpensive and hence appropriate for calculating properties of large systems without compromising accuracy. As an example, up to about 1000-atom electronic structure calculations \cite{Raza07} and up to about 150-atom transport calculations \cite{Raza08_prb} have been reported in silicon based systems with modest computational resources. In this paper, up to about 160 atoms calculations are presented. Contributions from five nearest neighbors are included. C-C atomic distance is taken as 1.44$\AA$, for which EHT parameters have been optimized. We find that incorporating about 3.5$\%$ decrease in C-C atomic distance \cite{Son06_PRL} for the edge carbon atoms results in a bandgap increase of about 52meV for N=18 W=2.1nm and bandgap decrease of about 64meV for N=19 W=2.2nm acGNR. Small bandgap changes are expected since wavefunctions for valence and conduction bands are deloaclized for acGNR. However, in zzGNR, these wavefunctions are localized on the edges and any atomic relaxation would have significant effect on the bandstructure. Since these variations are small in acGNR, we ignore any atomic relaxation in the reported calculations. Atomic visualization is done using GaussView \cite{GW03}.

\section{Electronic Structure} On a $p_z$ level of the tight-binding theory, two thirds of acGNRs are semiconducting with a bandgap inversely proportional to their widths and the other has zero bandgap depending on the chirality \cite{Nakada96}. However, one obtains a different result using the more sophisticated theory \cite{Son06_PRL}, such as EHT and DFT. First, the zero bandgap acGNRs also have a small bandgap that is inversely proportional to the width. Second, the remaining semicondcuting acGNRs only follow an inverse relation within its own category. For convenience, we propose to categorize them into $\alpha$-, $\beta$- and $\gamma$-acGNRs. This classification is similar to the ones used recently in Refs. \cite{Son06_PRL, Sahu08}. $\alpha$-acGNRs are N=8,11,14,... and have very small bandgap. $\beta$-acGNRs are N=9,12,15,... and $\gamma$ acGNRs are N=10,13,16,... acGNRs have also been classified into three subclasses in context of the orbital diamagnetism \cite{Wakabayashi99}. 

\begin{figure*}
\vspace{2.0in}
\hskip-3.0in\includegraphics{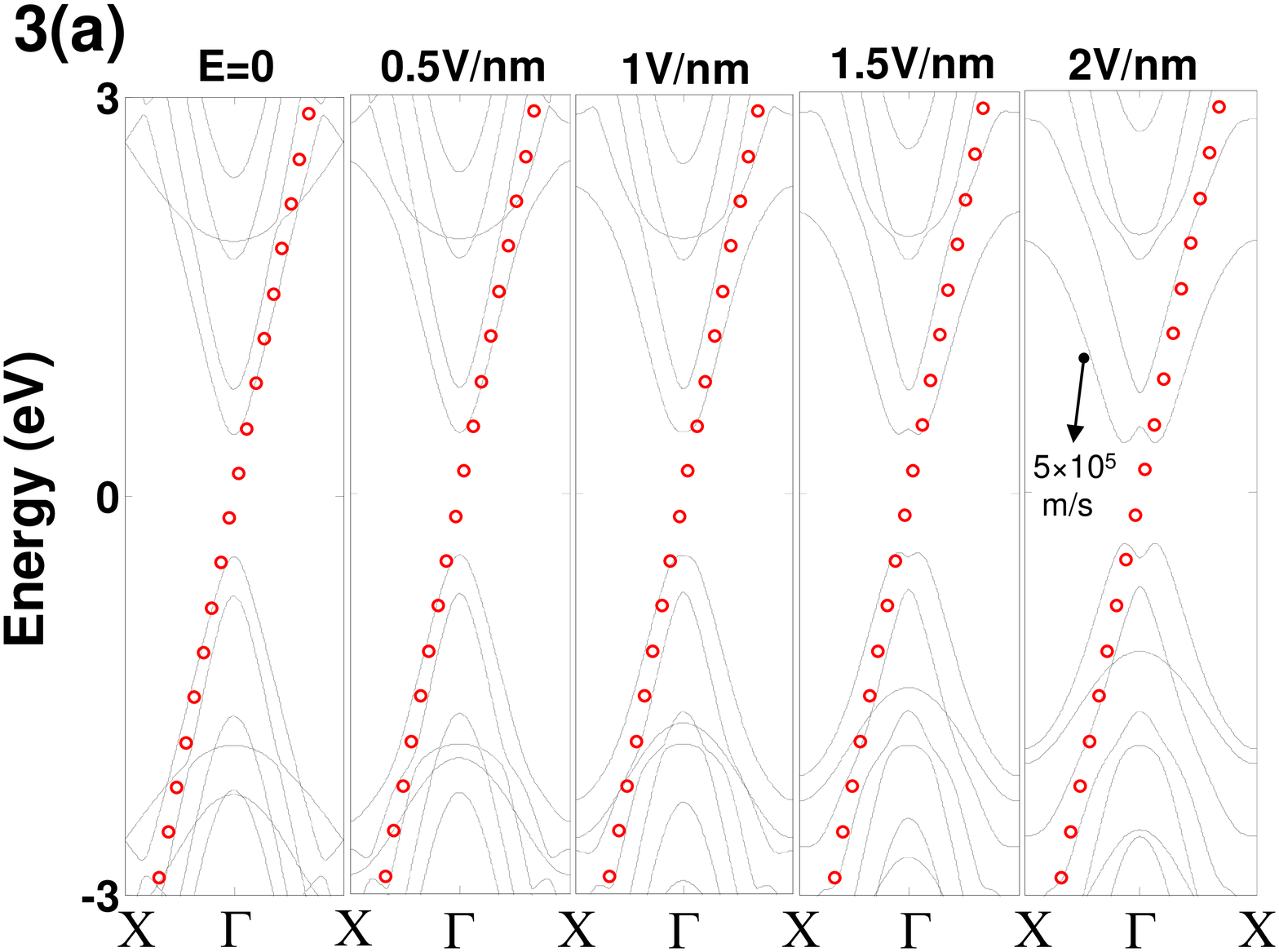}
\hskip3.6in\includegraphics{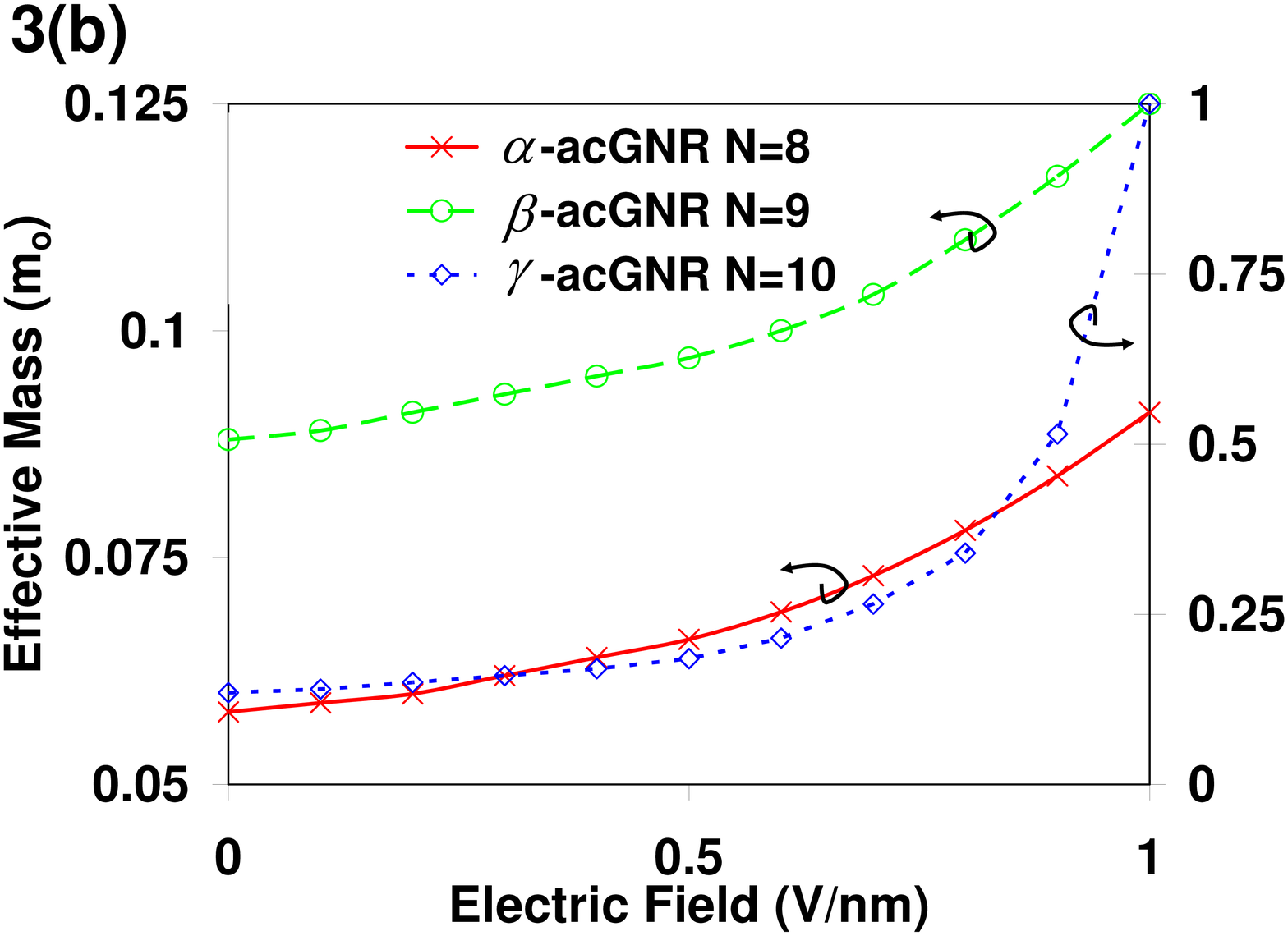}
\caption{(color online) Electric field modulation of band dispersions. (a) Variation of velocity in the width direction for N=10 $\gamma$-acGNR. The linear dispersion shown by red circles represents a value of $8.8\times 10^5$m/s - velocity around the Dirac point for graphene calculated using EHT. (b) Variation of effective masses. Effective masses are obtained by parabolic fits to the conduction bands within a few $k_BT$ of band edge for $\beta$-acGNRs and $\gamma$-acGNRs, and within a fraction of a $k_BT$ for $\alpha$-acGNRs.}
\end{figure*}

An electronic structure calculation for each type of acGNR is shown in Fig. 1. As can be seen that N=8 $\alpha$-acGNR has a small bandgap and has a nonlinear dispersion around the $\Gamma$ point. N=9 $\beta$-acGNR has a large bandgap with a parabolic dispersion around the $\Gamma$ point. Interestingly, N=10 $\gamma$-acGNR has a slightly larger bandgap with larger effective mass dispersion around the $\Gamma$ point and smaller velocity in the linear region away from the the $\Gamma$ point as compared to N=9 $\beta$-acGNR. We extract the bandgaps and effective masses within a few tens of meV around the band edges of these three types of acGNRs and plot them in Figs. 2(a) and (b), respectively. Fig. 2(a) is a computational verification of earlier results \cite{Son06_PRL} on a semi-empirical level. We find that incremental change in the bandgap of $\gamma$-acGNRs with respect to $\beta$-acGNRs is smaller in EHT than local density approximation of density functional theory \cite{Son06_PRL}. For each type of acGNR, bandgaps and effective masses are inversely proportional to the width with a different proportionality constant. The bandgap \textit{versus} width (\textit{W}) relations are given as:
\begin{eqnarray} E_{gap}=\begin{cases}{0.04eV/W(nm)\ for\ \alpha-acGNR\nonumber\cr 0.86eV/W(nm)\ for\ \beta-acGNR\nonumber\cr 1.04eV/W(nm)\ for\ \gamma-acGNR\nonumber}\end{cases} \nonumber\end{eqnarray} 
We find Fig. 2(b) important because some approaches toward graphene structures involve effective mass description \cite{Liang07}. Each type of acGNRs follow an inverse relation of effective mass with the width given below:
\begin{eqnarray} \frac{m}{m_o}=\begin{cases}{0.005/W(nm)\ for\ \alpha-acGNR\nonumber\cr 0.091/W(nm)\ for\ \beta-acGNR\nonumber\cr 0.160/W(nm)\ for\ \gamma-acGNR\nonumber}\end{cases} \nonumber\end{eqnarray} 
where $m_o$ is the free electron mass. It should be noted that using a $p_z$-orbital tight binding model, the effective mass follows the same inverse relation \textit{versus} width for all three types of acGNRs \cite{Liang07}. Furthermore, we determine the $p_z$-orbital tight binding parameters that reproduce the bandgaps as shown in Fig. 2(a). These parameters are 2.5eV and 2.7eV for $\beta$- and $\gamma$-acGNRs respectively. Since tight-binding parameter for $\gamma$-acGNRs is higher, we conclude that wavefunctions are hybridized more in this type of acGNR. This physical effect has some implications for electric field modulation as discussed in next section. 

\begin{figure*}
\vspace{2.0in}
\hskip-3.0in\includegraphics{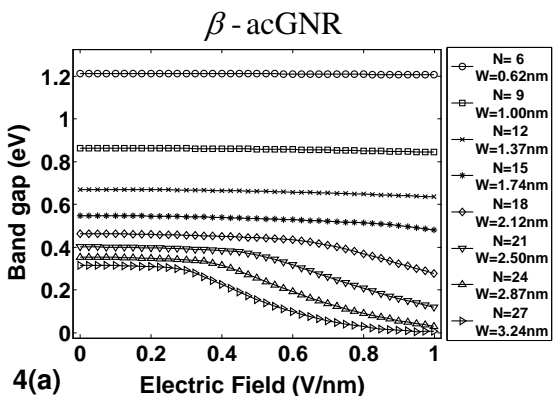}
\hskip3.6in\includegraphics{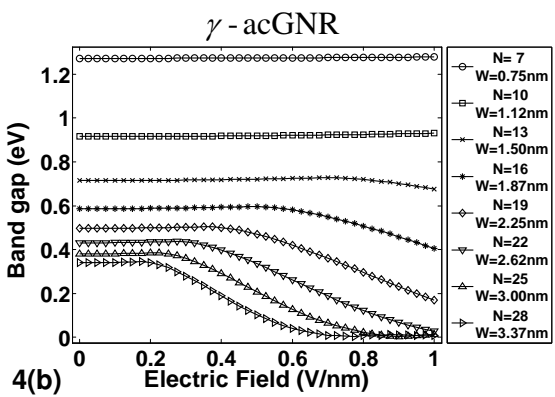}
\caption{Bandgap modulation. Bandgap as a function of width and electric field for (a) $\beta$-acGNRs and (b) $\gamma$-acGNRs. $\gamma$-acGNRs have larger bandgap modulation as compared to $\beta$-acGNRs.}
\end{figure*}

\section{Electric Field Modulation} 

Fig. 3(a) shows electric field modulation of the band structure for an N=10 $\gamma$-acGNR. The effective mass around the $\Gamma$-point increases with increasing electric field (E) and eventually changes sign, similar to Ref. \cite{Novikov07}. Furthermore, for E=0, the band dispersion in the linear regime away from the $\Gamma$-point shows velocity very close to the unconstrained graphene velocity (=$8.8\times10^5$ m/s) indicated by red (grey) circles. With increasing E, velocity in this linear regime away from the $\Gamma$-point decreases to about $5\times10^5$ m/s. In addition, the bandwidths of the valence and conduction bands are also decreasing. Moreover, a \textit{Mexican hat} structure is observable that has been seen in acGNR \cite{Novikov07}, carbon nanotubes \cite{Novikov06} and graphene bilayers \cite{Raza08, McCann06, Min07, Castro07}. These features are in qualitative agreement with the electric field effects reported in semiconducting acGNRs elsewhere using a continuum model \cite{Novikov07}. However, there are some quantiative differences which we address in this section. We show the extracted effective masses around the $\Gamma$-point for N=8, 9 and 10, which are $\alpha$-, $\beta$- and $\gamma$-acGNRs, respectively in Fig. 3(b). These effective masses are valid for tenths of $k_BT$ for $\alpha$-acGNRs and for a few $k_BT$ for $\beta$- and $\gamma$-acGNRs. After this energy scale, the band dispersions become linear again and remain so for about a few electron volts when they become nonlinear and hence saturate as shown in Fig. 3(a).  1V/nm electric field is within the dielectric breakdown limit of thermal SiO$_2$, which may result in higher electric field inside graphene due to smaller dielectric constant. Moreover, high-K dielectrics can be used to further enhance the electric field. However, such a high electric field may lead to dielectric reliability issues and is undesirable.

\begin{figure*}
\vspace{2.0in}
\hskip-3.0in\includegraphics{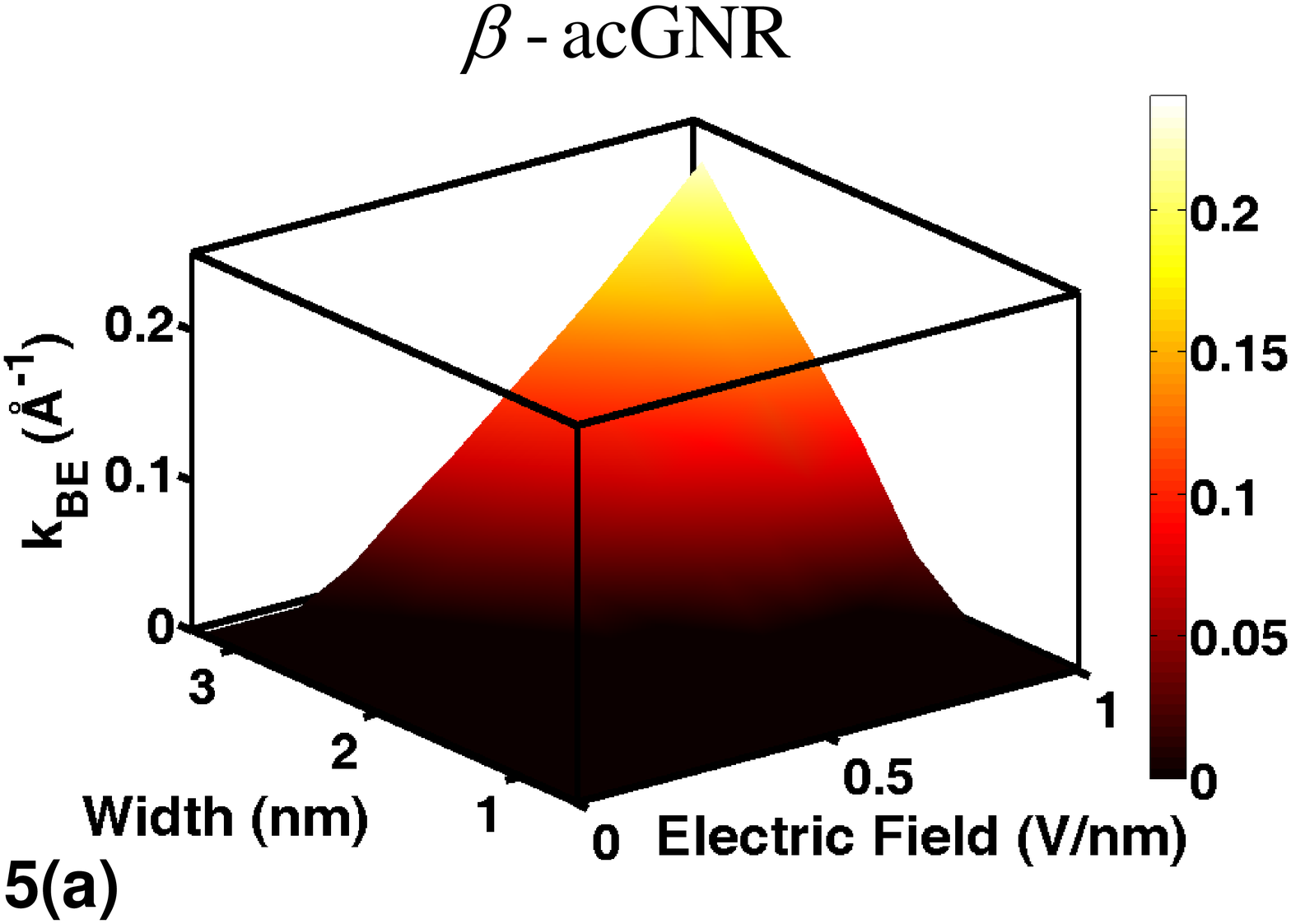}
\hskip3.6in\includegraphics{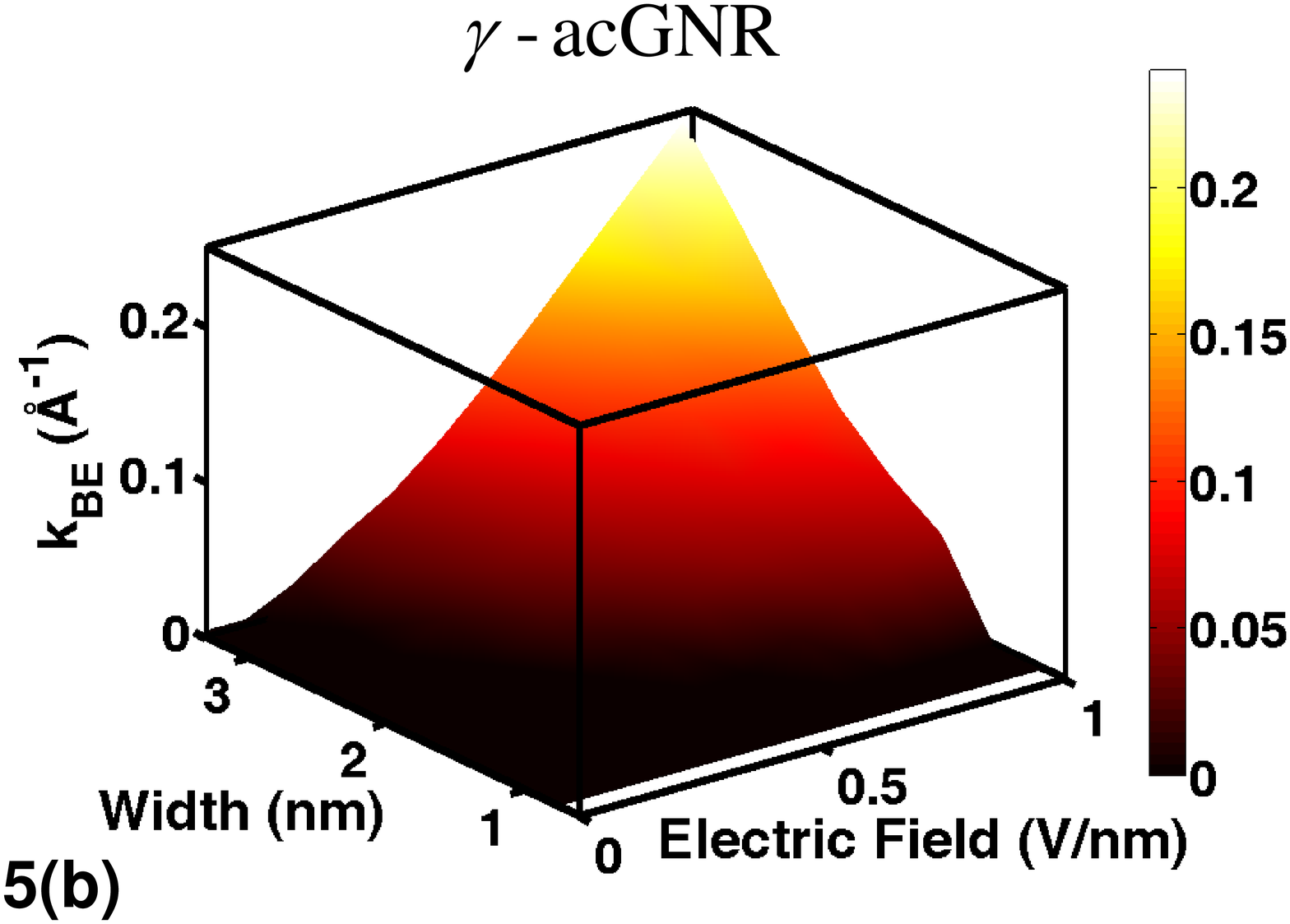}
\caption{(color online) The wavevector corresponding to band edge ($k_{BE}$) modulation. $k_{BE}$ as a function of width and electric field for (a) $\beta$-acGNRs and (b) $\gamma$-acGNRs. The value of k at X point is about 0.727$\AA^{-1}$. $\gamma$-acGNRs have larger shift in $k_{BE}$. }
\end{figure*}

In addition, the bandgap is modulated with increasing electric field. A clear feature is the location of wavevector corresponding to the conduction/valence band minimum/maximum. These two perturbations in the band structure are further explored in Figs. 4 and 5 respectively for $\beta$- and $\gamma$-acGNRs. In Fig. 4, we show bandgap modulation as a function of width and electric field. A threshold behavior is observed, similar to Ref. \cite{Novikov07}, where bandgap starts decreasing appreciably above a threshold electric field $E_t$. The dimensionless parameter $u_t=eE_t\times W/E_{gap}$ is reported as 4.5 for both kinds of acGNRs in Ref. \cite{Novikov07} using a continuum model. However, we find that this is different for these two acGNRs and is about 5.6 and 3.9 for $\beta$- and $\gamma$-acGNRs respectively. Moreover, for $\gamma$-acGNRs, bandgap decreases at a faster rate compared to $\beta$-acGNRs and thus $\gamma$-acGNRs have larger bandgap modulation. This is consistent because wavefunctions are more hybridized in $\gamma$-acGNRs and hence any perturbation affects the band structure more than $\beta$-acGNRs. 

Additionally, the bandgap for $\beta$-acGNRs monotonically decrease with electric field. However, the bandgap decreases appreciably only after a threshold electric field. This is different from Ref. \cite{Novikov07}, where below the threshold electric field, bandgap is constant and it decreses only after the threshold electric field. Furthermore, for $\gamma$-acGNRs, the bandgap first increases a little and then decreases - a feature although small, but not present in continuum calculations \cite{Novikov07}.  

With an appropriate electric field applied, one can reduce the bandgap of a semi-conducting acGNR to a few meV. We find that bandgap never becomes zero, whereas using a continuum model \cite{Novikov07}, one finds zero bandgap. In order to change the band structure, one has to incorporate perturbation on the order of the tight-binding parameter (2.5eV for $\beta$- and 2.7eV for $\gamma$-acGNRs). Therefore, an electric field of 1V/nm should not be able to induce a significant change in small width acGNRs due to small perturbation as shown in Fig. 4. However, the same electric field can change the electronic structure of a wider acGNRs due to larger potential variation. The physics behind this bandgap narrowing is the spectral shift of the conduction and valence band states on the two edges. This leads to downward and upward shift for conduction and valence band, respectively. Furthermore, in Fig. 5, we show the wavevector shift ($k_{BE}$) corresponding to conduction band minimum/valence band maximum. Again, $\gamma$-acGNRs have larger shift as compared to $\beta$-acGNRs. Overall, this shift can be as much as one third of the wavevector at X point. Unfortunately, we could not find a consistent set of $p_z$-orbital tight-binding parameters to reproduce Figs. 3, 4 and 5 simultaneously.

\section{Conclusions}

We have studied band structure and electric field modulation of acGNRs using EHT. The three types of acGNRs exhibit distinct electronic structure and electric field modulation properties. We extract important band structure parameters and a set of $p_z$-orbital tight-binding parameters benchmarked with extended H\"uckel theory to reproduce the bandgaps. Additionally, electric field modulation results are compared with a continuum model \cite{Novikov07}. We find that qualitative trends are the same, however there are some quantitative differences between the two models.  

The work is supported by National Science Foundation (NSF) and by Nanoelectronics Research Institute (NRI) through Center for Nanoscale Systems (CNS) at Cornell University. We are grateful to Tehseen Raza for GaussView \cite{GW03} visualizations and for reviewing the manuscript. We also thank D. S. Novikov for useful discussions.

\end{document}